\documentclass[english,11pt]{article}
\usepackage{amsmath,amsfonts,amssymb,amsbsy,amstext}
\usepackage[top=3cm, bottom=2cm, left=2cm, right=2cm]{geometry}
\usepackage{graphicx}
\usepackage[english]{babel}

\newcommand{\beq}{\begin{eqnarray}}
\newcommand{\eeq}{\end{eqnarray}}
\newcommand{\non}{\nonumber\\}

\begin{document}

\begin{titlepage}

\begin{center}
{\Large\bf Flavor of quiver-like realizations of effective supersymmetry}

\bigskip
{\large
Roberto Auzzi, Amit Giveon and Sven Bjarke Gudnason
}

\bigskip
{\it
Racah Institute of Physics, The Hebrew University,
 Jerusalem, 91904, Israel
}

\vskip 3cm

{\bf Abstract}
\end {center}
We present a class of supersymmetric models which address
the flavor puzzle and have an inverted hierarchy of sfermions.
Their construction involves quiver-like models with link fields
in generic representations.
The magnitude of Standard-Model parameters is obtained naturally
and a relatively heavy Higgs boson is allowed without fine tuning.
Collider signatures of such models are possibly within the reach of
LHC in the near future.

\vfill
\noindent
\rule{5cm}{0.5pt}\\
{\it\footnotesize
 auzzi(at)phys.huji.ac.il  \\
 giveon(at)phys.huji.ac.il  \\
 gudnason(at)phys.huji.ac.il
}

\end{titlepage}

\section{Introduction}

It is now an exciting period for supersymmetry (SUSY), as the LHC is
closing in on simplified SUSY models pointing theorists towards
certain parts of parameter space. As the experimental limits are
getting exceedingly harder for gauge mediated SUSY-breaking models which are
flavor blind (see \cite{arXiv:1110.6444} for a summary of current
collider constraints), one possibility
\cite{hep-ph/9507282,hep-ph/9607394} that is being extensively
explored now is ``effective supersymmetry.''
This ``more Minimal Supersymmetric Standard Model (MSSM)''
is supersymmetric in the UV but may differ significantly from
the MSSM; near the weak scale it necessarily includes just the
particles required for naturalness. Hence, this type of models
often possess an inverted hierarchy of sparticle masses,
i.e.~the stop being relatively light while the sup and scharm
may be considerably heavier (and similarly for the down-type squarks);
see
e.g.~\cite{arXiv:1004.2256,arXiv:1011.0730,arXiv:1105.2296,arXiv:1110.6670,arXiv:1110.6926,Delgado:2011kr}.

In addition to being motivated by current trends in collider limits,
the models of inverted hierarchy might be related to one of the
biggest and yet unsolved puzzles of particle physics, viz.~the flavor
problem. The first level of difficulty lies in providing an explanation
to the SM fermion mass hierarchy and the CKM matrix. The second level of
trouble is due to the mixing of squarks, generically giving
rise to flavor changing neutral currents (FCNCs). Several ways of
addressing the flavor puzzle have been put forward in the
literature. One of them involves horizontal symmetries which
suppress some of the Yukawa couplings \cite{CERN-TH-2519} and in turn
FCNCs. Another possibility is having a strongly
coupled conformal sector which provides large anomalous dimensions for
the first and second generations of sfermions
\cite{hep-ph/0006251,hep-ph/0104051,arXiv:1001.0637}.
Finally, a third possibility, investigated in this note, is that the
fermion textures are generated by irrelevant (gauge-invariant)
operators due to a quiver-like, UV completed theory, which in turn also
provides a sfermion hierarchy, i.e.~the sought-for inverted hierarchy
of sparticle masses. Our construction follows that of
\cite{CGK}, which uses bifundamental link fields in their quiver
construction, whereas we allow for generic representations of the link
fields.

The explicit model constructed in this note generates sfermion masses
via both gauge and gaugino mediated SUSY breaking; the first two
generations enjoy gauge-mediated contributions to their masses while
the third generation receives mass due to gaugino mediation.
In the examples that we study here, the Yukawa texture turns out to be
rather similar to that realized in single-sector SUSY-breaking models,
see
e.g.~\cite{hep-ph/9712389,hep-ph/9812290,arXiv:0704.3571,arXiv:0907.2689,arXiv:0911.2467},
where the first and second generation sfermions are composite while the
third generation is elementary.
In terms of the messenger scale $M$ and the Higgsing scale of the link
field $v$, there are in principle three possible regimes to explore:
$M\ll v$, i.e.~providing a flavor blind sparticle spectrum; $M\sim v$,
i.e.~giving rise to a relatively mild sparticle mass hierarchy; and finally, $M\gg
v$, which potentially provides a large hierarchy.~\footnote{The latter
gives rise to Landau poles in our examples of interest (even for $M$ of order of
$100v$ or so).}

While restricting here to a minimal gauge mediation (MGM) sector of
SUSY breaking, this can be extended using the General Gauge
Mediation (GGM) formalism
\cite{arXiv:0801.3278,arXiv:0901.1326,arXiv:1003.2661}.
For instance,
to realize our setup in a dynamical SUSY-breaking model, as
e.g.~\cite{arXiv:1008.2215}, one needs to consider a more general
messenger sector. Such an embedding of direct gaugino
mediation was studied in \cite{arXiv:1107.1414}.

The organization of this note is as follows. In sec.~\ref{sec:2nodes},
we present the minimal version of our models,
based on a quiver-like construction with two gauge groups;
this realizes the observed mass
hierarchy between the third and the first two generations of
fermions. In sec.~\ref{sec:higgs}, we consider
naturalness in our class of models,
even for a relatively heavy Higgs particle.
In sec.~\ref{sec:3nodes}, we describe an extension
which gives rise also to the
hierarchy between the first and second generations of fermions.
Finally, we conclude in
sec.~\ref{sec:discussion} with a short discussion and outlook.

\section{Two nodes model} \label{sec:2nodes}

\begin{figure}[!ht]
\begin{center}
\includegraphics[width=0.35\linewidth]{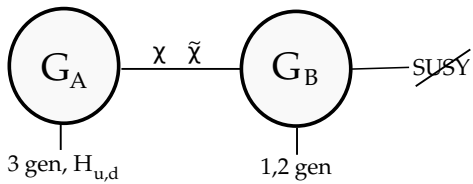}
\caption{A diagram describing the model with $G_A,G_B$ being gauge
  groups and $\chi,\tilde{\chi}$ being link fields. }
\label{fig:quiver2nodes}
\end{center}
\end{figure}

Gaugino mediation \cite{hep-ph/9911293,hep-ph/9911323}
produces a spectrum where the sfermion
masses are suppressed with respect to the gaugino masses.
%can be obtained by
%can be realized as
It can be deconstructed \cite{hep-ph/0106044,hep-ph/0106098} in terms of
quiver gauge theories Higgsed to the
Standard Model (SM) gauge group by the link fields at low energy.
In this setting all the SM matter fields are charged under the same gauge
group which is connected by link fields (directly or via other gauge
groups) to another gauge group under which the messengers of SUSY breaking
are charged.
In order to produce the sought-for inverted hierarchy of sfermion
masses, the supersymmetric SM generations are split such that the
first two generations are charged with respect to the same group as
the messenger fields while the third generation as well as the Higgses
are charged under a different group.
In this section we consider the two nodes model in fig.~\ref{fig:quiver2nodes}.

Let us step back for a moment and consider the following
representations:~\footnote{They consist of the representations
  identical to those of a single SM generation.}
\begin{center}
\begin{tabular}{c|c|c|c}
R & $SU(3)$ & $SU(2)$ & $U(1)$ \\
\hline
$Q$ & {\bf 3} & {\bf 2} & $1/6$ \\
$u^c$ & $\bar{\bf 3}$ & {\bf 1} & $-2/3$ \\
$d^c$ & $\bar{\bf 3}$ & {\bf 1} & $1/3$ \\
$L$ & {\bf 1} & {\bf 2} & $-1/2$ \\
$e^c$ & {\bf 1} & {\bf 1} & $1$
\end{tabular}
\end{center}
Without any prejudices we can now consider couples of link fields
$\chi_R,\tilde{\chi}_R$, in the representation $(R,\bar{R})$
and $(\bar{R},R)$ of the group $G_A\times G_B$,
and we choose $G_A,G_B=SU(3)\times SU(2)\times U(1)$, for simplicity.
$R$ can be one of the representations given in the table above.
A single couple of link fields is in general not sufficient for
providing all the SSM Yukawa couplings and hence we propose the
skeletal link of fig.~\ref{fig:quiver2nodes} to consist of at least
two couples of fields e.g.~$\{\chi_{R_1},\chi_{R_2},\cdots\}$. A
single couple of bifundamental link fields of $SU(5)$ corresponds in
our notation to $\{\chi_{d^c},\chi_L\}$, which is the case studied in
detail in \cite{CGK}, while \cite{arXiv:1105.2296} considered the case
of $\{\chi_Q,\chi_L\}$.

When a choice is made, the link fields give rise to Yukawa textures
for the fermions of the SM in terms of higher dimension operators. As
an example, we can have
\beq
\frac{\lambda^u_{ij}}{\Lambda^2} Q_i H_u u_j^c \chi_{u^c} \chi_Q \, ,
\label{qhuxx}
\eeq
where $i,j=1,2$ are generation indices and the labels on the $\chi$s
denote the representation under which they transform. This particular
example generates a Yukawa matrix
\beq
Y^u_{ij} = \lambda^u_{ij} \epsilon_u \epsilon_Q \, , \qquad i,j=1,2 \,
. \label{eq:Yukawa_ex}
\eeq
Here we have defined $\epsilon_R\equiv v_{R}/\Lambda$, where
$v_{R}$ is the VEV of $\chi_R,\tilde{\chi}_R$ and $\Lambda$
is the UV scale of flavor dynamics.
Similar operators are needed for the complete
set of three generations, as well as for the down and lepton sectors.

To break $G_A\times G_B$ to the SM group $G_{\rm SM}$
we must include link fields charged under both $SU(3)$ and $SU(2)$.
Moreover, to choose the ideal representations for the link fields we
need to check if they reproduce naturally the quark masses
and the CKM matrix.
The minimal choice required for these purposes is one of the
following five possibilities:
$\{\chi_L,\chi_{u^c}\}$, $\{\chi_L,\chi_{d^c}\}$,
$\{\chi_Q,\chi_L\}$, $\{\chi_Q,\chi_{u^c}\}$,
$\{\chi_Q,\chi_{d^c}\}$.
It turns out that one can easily obtain
such textures with all $\lambda^u_{ij},\lambda^d_{ij}$
generically being close to one, in the cases
$\{\chi_Q,\chi_{u^c}\}$ and $\{\chi_Q,\chi_{d^c}\}$.
The case $\{\chi_Q,\chi_L\}$ requires an extra hierarchy
of a factor of roughly 3 between the different couplings
relative to the previous ones,
while the choices $\{\chi_L,\chi_{u^c}\}$, $\{\chi_L,\chi_{d^c}\}$
require an extra factor of 20 tuning, instead.

Interestingly, the ${\bf 10}$ of $SU(5)$ decomposes under $SU(3)\times SU(2)\times
U(1)$ as $Q\oplus u^c\oplus e^c$.
The texture of the quark sector is blind to the inclusion of the
link $\chi_{e^c}$ and hence does not alter the above argument. The
effect on the sleptonic sector is a possible increase in the slepton
masses which can be useful for the RG evolution not to drive the stau
tachyonic.
In the following we will thus focus on the choice of link fields
$\{ \chi_Q, \chi_{u^c},  \chi_{e^c} \}$.

Since we include higher dimension operators,
B or L-violating operators, such as
\beq
\frac{Q Q Q L}{\Lambda} \, , \qquad \frac{u^c u^c d^c e^c}{ \Lambda}
\, , \qquad
\frac{L H_u L H_u}{\Lambda} \, , \label{lhlh}
\eeq
cannot be avoided by imposing R-parity. If
$\Lambda$ is below the GUT scale, we are forced to impose either baryon
or lepton number conservation to prevent proton decay. The last
operator in eq.~\eqref{lhlh} could be envisioned to provide neutrino
masses via a
seesaw mechanism if $\Lambda$ is bigger than roughly
$10^{13}$ GeV. In the following we will usually impose both R-parity as well
as baryon and lepton number conservation.

\subsection{Two nodes model with link fields $\{ \chi_Q, \chi_{u^c},
  \chi_{e^c} \}$}

Due to the discussion above, we consider
the link fields $\{\chi_Q,\chi_{u^c},\chi_{e^c}\}$.
The Yukawa textures in this case are
\begin{align}
Y^u \sim
\begin{pmatrix}
\epsilon_u \epsilon_Q &\epsilon_u \epsilon_Q  & \epsilon_Q \\
\epsilon_u \epsilon_Q & \epsilon_u \epsilon_Q  & \epsilon_Q  \\
 \epsilon_u & \epsilon_u & 1
\end{pmatrix} \, , \qquad
Y^d \sim
\begin{pmatrix}
\epsilon_u \epsilon_Q &\epsilon_u \epsilon_Q  & \epsilon_Q \\
\epsilon_u \epsilon_Q & \epsilon_u \epsilon_Q  & \epsilon_Q  \\
\epsilon_Q^2 & \epsilon_Q^2 & 1
\end{pmatrix}  \, , \qquad
Y^e \sim
\begin{pmatrix}
\epsilon_u \epsilon_Q &\epsilon_u \epsilon_Q & \epsilon_u \epsilon_Q \\
\epsilon_u \epsilon_Q & \epsilon_u \epsilon_Q &\epsilon_u \epsilon_Q \\
\epsilon_e & \epsilon_e & 1
\end{pmatrix} \, , \label{eq:yukawas}
\end{align}
where each element of the matrices is multiplied by the coefficient
$\lambda^{u,d,e}_{ij}$, respectively.\footnote{The texture of $Y^u$ is
  identical to that of the single-sector SUSY breaking in
  \cite{hep-ph/9712389,arXiv:0907.2689}, while $Y^d$ differs slightly
  in the lower triangular part.}
For randomly chosen order one coefficients, generically this texture
predicts the following ratios
\beq
\frac{m_u}{m_t} \sim  \frac{m_c}{m_t} \sim  \frac{m_d}{m_b} \sim  \frac{m_s}{m_b}
\sim  \frac{m_e}{m_\tau}  \sim  \frac{m_\mu}{m_\tau} \sim
\mathcal{O}(\epsilon_u \epsilon_Q) \, ,\qquad
\frac{m_t}{m_b}\sim \frac{m_t}{m_\tau} \sim \tan\beta\, ,
\eeq
explaining only the mass hierarchy between the second and third
generations, whereas the hierarchy between the first and second
generations can be reproduced by tuning the coefficients in such a
way that the first two rows are nearly parallel vectors.
The optimal value of the $\epsilon$s can be read off the mass ratio
$m_s/m_b$ to be roughly $\epsilon_Q\simeq \epsilon_u\simeq 1/10$.
With these values, the mass hierarchy in each sector --
the up-type quarks, down-type quarks and leptons --
is reasonable and, moreover,
for a sufficiently large $\tan\beta$, the ratio between
the down-type quark and lepton masses to the up-type masses
is also reasonable.

Note that by not imposing lepton number conservation at the scale $\Lambda$,
a neutrino mass matrix is also generated (in flavor basis):
\beq
m_\nu \sim   \frac{v_{\rm ew}^2}{\Lambda}
\begin{pmatrix}
\epsilon_u^2 \epsilon_Q^2 & \epsilon_u^2 \epsilon_Q^2 & \epsilon_u \epsilon_Q \\
\epsilon_u^2 \epsilon_Q^2 & \epsilon_u^2 \epsilon_Q^2 &\epsilon_u \epsilon_Q \\
\epsilon_u \epsilon_Q & \epsilon_u \epsilon_Q & 1
\end{pmatrix} \, , \qquad v_{\rm ew} = 174 \, {\rm GeV} \, .
\label{eq:mneutrinos}
\eeq
Even for $\Lambda \sim 10^{14} \, {\rm
  GeV}$, which potentially could produce viable neutrino masses, the
matrix $m_\nu$ requires some tuning to reduce the big hierarchy in the
masses and improve the lack of large mixing angles.

The link fields give rise to the masses of the SM fermions,
via $\Lambda$-suppressed terms, but we also
need to ensure that they decouple from the SM,
ideally by a similar effect.
To avoid Landau poles
they should have masses at least of the order of their VEVs,
$\langle\chi_R\rangle=\langle\tilde{\chi}_R\rangle=v_R$.
Assuming that the superpotential for the link fields is indeed
generated by physics at the UV scale $\Lambda$, it takes the form
\begin{align}
{\cal W} = &
-\mu_Q \, \chi^{\alpha i}_{\beta j} \tilde{\chi}^{\beta j}_{\alpha i}
-\mu_u \, \chi^{\alpha}_{\beta} \tilde{\chi}^{\beta}_{\alpha}
-\mu_e \, \chi \tilde{\chi} \non &
+\frac{\chi^{\alpha i}_{\beta j} \tilde{\chi}^{\beta j}_{\gamma s}
\chi^{\gamma s}_{\sigma r} \tilde{\chi}^{\sigma r}_{\alpha i}}{\Lambda}
+\frac{\chi^{\alpha i}_{\beta j} \tilde{\chi}^{\beta r}_{\alpha t}
\chi^{\gamma t}_{\sigma r} \tilde{\chi}^{\sigma j}_{\gamma i}}{\Lambda}
+\frac{\chi^{\alpha i}_{\beta j} \tilde{\chi}^{\sigma j}_{\gamma i}
\chi^\gamma_\sigma \tilde{\chi}^\beta_\alpha  }{\Lambda}
+\frac{\chi^{\alpha}_{\beta}\tilde{\chi}^\beta_{\gamma}\chi^{\gamma}_{\sigma}
\tilde{\chi}^\sigma_{\alpha}  }{\Lambda}
+\frac{(\chi \tilde{\chi})^2}{\Lambda} \, , \label{superpote}
\end{align}
where the field $\chi^{\alpha i}_{\beta j}$ transforms as $Q$,
with $\alpha$ the $SU(3)$ index and $i$ the
$SU(2)$ index under the group $G_A$, and as $\bar{Q}$ for $\beta,j$ under
the group $G_B$. The field $\chi^\alpha_\beta$ transforms as $\bar{u^c}$
under the group $G_A$ and as $u^c$ under the
group  $G_B$. Finally, $\chi$ is a singlet
with charge $-1$ under $U(1)_A$ and $1$ under $U(1)_B$. The fields
with tildes transform under the conjugate representation
with respect to the fields without tildes.
This is not the most general superpotential,
but it is sufficient to break $G_A\times G_B$ to $G_{\rm SM}$
as well as to give mass to all the link field components which are
not eaten by the super-Higgs mechanism. We take the coefficients
of the irrelevant terms in the above superpotential to be equal as an
illustrative example. At the minimum where $G_A\times G_B\to G_{\rm SM}$,
we have
\beq
\langle\chi^{\alpha i}_{\beta j}\rangle =
v_Q\delta^\alpha_\beta\delta^i_j \, , \qquad
\langle\chi^\alpha_\beta\rangle =
v_u\delta^\alpha_\beta \, , \qquad
\langle\chi\rangle = v_e \, ,
\eeq
where the VEVs $v_R$ are related to $\mu_R$ in eq.~\eqref{superpote} via
\begin{align}
\mu_Q = \frac{14v_Q^2 + 3v_u^2}{\Lambda} \, , \qquad
\mu_u = \frac{6v_Q^2 + 2v_u^2}{\Lambda} \, , \qquad
\mu_e = \frac{v_e^2}{\Lambda} \, .
\end{align}
The spectrum obtained at this minimum of the potential, for equal VEVs,
is as follows. An (${\bf 8},{\bf 1}$) and a (${\bf 1},{\bf 3}$) have masses
$4\epsilon v$; two (${\bf 8},{\bf 1}$)'s have masses
$5\epsilon v$ and $9\epsilon v$, respectively;
two (${\bf 8},{\bf 3}$)'s have masses $11\epsilon v$ and $15\epsilon
v$, respectively; and, finally, four singlets have masses
$\{1.3\epsilon v,4\epsilon v,4\epsilon v,30.7\epsilon v\}$. Since we
take $\epsilon \simeq 1/10$, most of the matter acquires a mass of
order $v$.
There are still remaining an (${\bf 8},{\bf 1}$), a (${\bf 1},{\bf
  3}$) and one singlet, which acquire a mass via the super-Higgs
mechanism. In addition, there is a massless Goldstone boson coming from
a global $U(1)$ symmetry under which $\chi$ has charge $1$ and
$\tilde{\chi}$ has charge $-1$. We can get rid of this state by
introducing e.g.~the following term in the superpotential:
$\epsilon_{\alpha_1 \alpha_2 \alpha_3} \epsilon^{\beta_1 \beta_2 \beta_3}
 \chi^{\alpha_1}_{\beta_1}  \chi^{\alpha_2}_{\beta_2}  \chi^{\alpha_3}_{\beta_3} \chi \chi/\Lambda^2$.

\subsection{RG evolution}

Let us first discuss the various scales in the problem. We
found that the $\epsilon$s should be of order $1/10$,
which means that $\Lambda$ is only an order of magnitude bigger than $v$.
The mass-squared matrix of the gauge bosons is
\begin{align}
\mathcal{M}^2_{V_k} &= 2 v_{k}^2
\begin{pmatrix}
g_{A_k}^2 & -g_{A_k} g_{B_k} \\
-g_{A_k} g_{B_k} & g_{B_k}^2
\end{pmatrix}  \, , \\
v_{1}^2 &\equiv \frac{v_{Q}^2}{5} + \frac{8v_{u}^2}{5}
  + \frac{6v_{e}^2}{5} \, , \qquad
v_{2}^2 \equiv 3v_{Q}^2  \, , \qquad
v_{3}^2 \equiv 2v_{Q}^2 + v_{u}^2  \, . \label{vevvini}
\end{align}
Its zero eigenvalues amount to the SM gauge particles, while
the heavy gauge bosons have masses
\beq
m_{v_k}^2= 2\left(g_{A_k}^2+ g_{B_k}^2\right) v_{k}^2 \, ,
\eeq
where $k=1,2,3$ denotes the gauge groups $U(1)$, $SU(2)$ and $SU(3)$,
respectively.
A comment about the group theoretic data specifying our choice of
representations of the link fields is in store, viz.~it is encoded in
the coefficients of the VEVs in eq.~\eqref{vevvini}.
At the Higgsing scale $m_v$, which we approximate by $3v$,
the SM gauge couplings are given in terms of the
gauge couplings $g_{A,B}$ by
\beq
\frac{1}{g_k^2} = \frac{1}{g_{A_k}^2} + \frac{1}{g_{B_k}^2} \ .
\eeq
Assuming for simplicity that the components of the link fields all
acquire a mass of order $v$, then below the scale $v$ the
running of the gauge couplings $g_k$ is given by that of
the MSSM, i.e.
\beq
\frac{d \alpha^{-1}_k}{d t} = -\frac{b_k}{2 \pi} \, , \qquad
b_1 = \frac{33}{5} \, , \qquad b_2 = 1 \, , \qquad b_3 = -3 \, .
\eeq
Now, from the scale $v$ up to $m_v\sim 3v$, the
$\beta$-function coefficients become
\beq
b_1 = \frac{33}{5} \, , \qquad b_2 = 35 \, , \qquad b_3 = 24 \, ,
\label{modifiche}
\eeq
where the contribution to $b_3$ corresponds to $9$ adjoints, whereas
to $b_2$ it is analogous to $17$ adjoints. This change in the
$\beta$-functions is rather drastic; for instance, a factor of $3$ running in
energy decreases $\alpha_{2,3}^{-1}$ by  $6$ and $4$,
respectively.

Above the Higgsing scale, the $\beta$-function coefficients are split
into two sets, corresponding to the two gauge groups depicted in
fig.~\ref{fig:quiver2nodes}, and read
\beq
b_{A_1}=\frac{49}{5} \, , \qquad b_{A_2}=15 \, , \qquad b_{A_3}=8 \, ,
\qquad
b_{B_1}=\frac{56}{5} \, , \qquad b_{B_2}=16 \, , \qquad b_{B_3}=10 \, .
\eeq
The contribution from the link fields is $36/5$ to $b_1$, $18$ to
$b_2$ and $15$ to $b_3$.
These large values of the $\beta$-function coefficients result in the
fact that the gauge couplings run into Landau poles rather
quickly. Hence, the region of parameter space corresponding to weakly
coupled gaugino mediation ($M\gg m_v$) is somewhat far
fetched.\footnote{Taking for instance $M\sim 10^{15}$ GeV
  ($M\sim 10^5$ GeV), and assuming that the link fields have a mass $v$,
  a Landau pole arises near $120 m_v$ ($6m_v$). Assuming instead that
  the link fields have a mass $m_v\sim 3v$, the Landau pole moves up
  to $400 m_v$ ($20m_v$). Raising the mass of all the link field
  components to $m_v$ requires somewhat ``large'' coefficients of the
  irrelevant terms in the superpotential \eqref{superpote}.}
Models which do not contain the link field $\chi_Q$ do not
suffer as severely from fast running as is the case here.

As already mentioned, two other regimes are possible in our model,
viz.~$M\sim m_v$ and $M\ll m_v$. The case $M\sim m_v$, which is our
main interest, provides a mild hierarchy of sparticle masses, i.e.~the
first two generations of squarks acquire a mass which is a factor of a
few larger than that of the third generation.
The other case,
$M\ll m_v$, gives rise to a gauge mediation sparticle spectrum,
which is nearly flavor blind, and hence flavor constraints are satisfied
trivially. However, recent collider limits place more restrictive bounds in this case.
Finally, the scale $m_v$ can be anywhere between about $10^5$ GeV and
the Planck scale, but if we insist on gauge coupling unification it should
be placed near the GUT scale.

\subsection{Sparticle spectrum}

The sfermion masses in the type of setup discussed above have been studied in detail in
\cite{arXiv:1009.0012,arXiv:1009.1714,arXiv:1009.2086,arXiv:1101.5158,Auzzi:2011wt}.
For simplicity, we restrict to a minimal messenger sector realized by coupling an
F-term spurion $S$ to a single pair of messengers, $T,\tilde{T}$,
in the ${\bf 5}$ and $\bar{\bf 5}$ of $SU(5)$, via
\beq
\mathcal{W}_T = S T \tilde{T} \, , \qquad
\langle S \rangle = M + \theta^2 F \, .
\eeq
We introduce the variables
\beq
x \equiv \frac{F}{M^2} \ , \qquad
y_k \equiv \frac{m_{v_k}}{M} \ , \qquad k=1,2,3 \ .
\label{eq:variables}
\eeq
The gaugino masses are given by those of minimal gauge mediation
\cite{hep-ph/9608224}:
\beq
m_{\tilde{g}_k}=\frac{\alpha_k}{4 \pi} \frac{F}{M} n_k \, q(x) \, ,  \qquad
q(x)=\frac{1}{x^2}\left[(1+x ) \log(1+x) + (1-x) \log(1-x) \right] \, ,
\label{mgauginos}
\eeq
where $\alpha_{k}^{-1} = 4\pi/g_{A_k}^2 + 4\pi/g_{B_k}^2$
and $n_k$ is the Dynkin index of the messenger field.

The sfermion masses are given in eq.~(4.2) of \cite{Auzzi:2011wt},
\beq
m_{\tilde{f}_l}^2= 2 \sum_{k=1}^{3}
\left(\frac{\alpha_{k}}{4\pi}\right)^2
\left(\frac{F}{M}\right)^2  \,
 C_{2 k}^{\tilde{f}} \,
n_k \mathcal{E}^l(x,y_k) \ , \label{sferm2}
\eeq
where $C_{2 k}^{\tilde{f}}$ is the quadratic
Casimir of the representation under which the sfermion ${\tilde f}$ transforms,
while the index $l$ runs over generations.\footnote{The
  expression \eqref{sferm2} is valid for link
  fields in {\it any} representation $(R,\bar{R})$, and the effect of the
  representation is encoded in the gauge particle masses $m_{v_k}$.}
The function $\mathcal{E}^l$ for the first and second generation is
given by
\begin{align}
\mathcal{E}^{1,2}(x,y,\lambda_2) &= \frac{1}{x^2}\bigg[
\alpha_0(x)
-\left(1-\lambda_2^2\right)\alpha_1(x,y)
-(1-\lambda_2)^2 y^2 \alpha_2(x,y)
-\frac{2(1-\lambda_2)}{y^2}\beta_{-1}(x)
+ \beta_0(x) \non &
\phantom{=\frac{1}{x^2}\bigg[\ }
+\frac{2(1-\lambda_2)}{y^2}\beta_1(x,y)
+(1-\lambda_2)^2\beta_2(x,y) \bigg] \, , \qquad
\lambda_2\equiv \frac{g^2_{B_k}}{g^2_k} \, ,
\end{align}
whereas for the third generation it reads
\beq
\mathcal{E}^3(x,y) = \frac{1}{x^2} \left[
\alpha_0(x) - \alpha_1(x,y) - y^2 \alpha_2(x,y)
-\frac{2}{y^2}\beta_{-1}(x) + \beta_0(x) + \frac{2}{y^2} \beta_1(x,y)
+\beta_2(x,y) \right] \, .
\eeq
The $\alpha$s and $\beta$s
are defined in Appendix A of \cite{Auzzi:2011wt}.
The soft mass of the link field is also given by eq.~\eqref{sferm2},
with $\mathcal{E}^{\rm link}=\mathcal{E}^1+ \mathcal{E}^3$ and an
appropriate quadratic Casimir (see \cite{Auzzi:2011wt} for details).

Finally, we present an example of the spectrum for low-scale
as well as for high-scale mediation in table \ref{masses2nodes}.
Here we have assumed that the mass of all the link fields is near
$v$ and we have used eq.~\eqref{modifiche} for the gauge couplings as well
as two-loop corrections to the sfermion masses coming from the link
field soft masses \cite{hep-ph/9311340}. The RG evolution from the
scale $v$ down to the weak scale and the determination of the pole masses
were done using SOFTSUSY \cite{hep-ph/0104145}.

\begin{table*}[!htp]
\begin{center}
\begin{tabular}  {| l | l | l | } \hline
 &  $M=5 \times 10^{5}$  &  $M= 10^{15}$ \\
$\tan\beta=20$ &   $F/M=1.6 \times 10^5$  &   $F/M=1.6 \times 10^5$ \\
\hline
$(\alpha^{-1}_{B_1},\alpha^{-1}_{B_2},\alpha^{-1}_{B_3})$   & $(30,13,5)$  & $(10,10,10)$   \\
$(y_1,y_2,y_3)$   & $(1.4,1.9,2.8)$    & $(1.9,2.0,2.1)$   \\
$(v_{Q},v_{u},v_{e})$   & $(0.55 M,0.55 M,0.55 M)$  & $(0.55 M,0.55 M,0.55 M)$   \\
\hline
$m_{\tilde{g}}$    & $1287$ & $1218$    \\
$m_{\tilde{\chi}_0}$   &$(217,416,569,590)$ &$(209,407,681,692)$  \\
$m_{\tilde{\chi}_{\pm}}$    & $(416,589)$  & $(406,692)$    \\
\hline
$(m_{\tilde{u}_L}, m_{\tilde{d}_L},m_{\tilde{u}_R},m_{\tilde{d}_R})$
&$(3241,3242, 3094, 3085)$ &$(2279,2280, 2024, 1925)$   \\
$(m_{\tilde{t}_1},m_{\tilde{t}_2},m_{\tilde{b}_1},m_{\tilde{b}_2})$
& $(1114, 1218, 1195, 1222)$  &  $(833, 1063, 1018, 1051)$   \\
$(m_{\tilde{e}_R}, m_{\tilde{e}_L},m_{\tilde{\nu}_e})$
& $(420, 1005, 1002)$  & $(1088, 1373, 1370)$    \\
$(m_{\tilde{\tau}_1},m_{\tilde{\tau}_2},m_{\tilde{\nu}_\tau})$ &
 $(111, 309, 289)$ &  $(260, 468, 466)$ \\
\hline
    $(m_{h_0},m_{H_0},m_{A_0},m_{H_{\pm}})$
      &$(115, 604, 604, 609)$   &$(115, 775, 775, 779)$\\
  \hline
 $\mu$  &$581$  & $694$ \\
 \hline
\end{tabular}
\caption{Sparticle pole masses in units of GeV in
  numerical examples, for $\tan\beta=20$ and a single
  messenger. $\epsilon_{Q,u}$ are taken to be $0.1$. }
\label{masses2nodes}
\end{center}
\end{table*}

\subsection{Flavor constraints}

So far, we have constructed a model which gives rise to an inverted hierarchy of
squark masses and chose link fields which produce the
measured quark masses and the observed CKM matrix naturally. Yet,
we should check if the model satisfies the current flavor constraints.
The most stringent constraints are due to CP-violating FCNCs, implying
that a couple of complex phases must be rather small -- at the percent level.
Constraints from meson oscillations are somewhat easier to satisfy since
our model has degenerate sfermion masses for the first two generations,
which ameliorate the danger of unacceptable $K-\bar{K}$ and $D-\bar{D}$ mixing.
Constraints due to $B_d-\bar{B}_d$ mixing turn out to be the most important
in this model; less important are those due to $B_s-\bar{B}_s$ mixing.
Next, we establish that in a rather large regime of parameter
space, the meson mixing constraints are satisfied, as a result of
the sparticle mass hierarchy as well as quark-squark alignment.

Let us first define the fermion mass matrices $m_u=Y^u v_u$,
$m_d=Y^d v_d$, $m_e=Y^e v_d$. We can rotate to the mass eigenstate
basis via
\beq
(U_L^u)^\dagger m_u U_R^u = D^u \, , \qquad
(U_L^d)^\dagger m_d U_R^d = D^d \, ,
\eeq
in terms of which the CKM matrix is $V_{\rm CKM}=(U_L^u)^\dag U_L^d$.
The most general Yukawa matrices which give correct quark masses as
well as the $V_{\rm CKM}$ can be written as
\beq
Y^u v_u = A V_{\rm CKM}^\dag D^u B \, , \qquad
Y^d v_d = A D^d C \, ,\qquad
v_{u,d}\equiv \langle H^0_{u,d}\rangle \, ,\qquad
\tan\beta\equiv\frac{v_u}{v_d} \, ,
\label{rotazioni}
\eeq
with $A$, $B$ and $C$ being arbitrary $SU(3)$ matrices. Hence, it
follows that by ignoring complex (CP) phases we are dealing with three
3-spheres of parameter space. We find $A,B,C$ in such a way that the
coefficients $\lambda^{u,d}_{ij}$ (see eq.~\eqref{eq:Yukawa_ex}) are
all in a certain range, say $[0.11,1.1]$. There are a lot of such
similar solutions, all giving rise to the measured fermion masses
and CKM matrix but not necessarily to the same
flavor constraints, which we shall analyze next.

Consider the sfermion mass-squared matrix
\beq
\mathcal{M}^2_{\tilde{f}}=
\begin{pmatrix}
(\mathcal{M}^2_{\tilde{f}})_{LL} &  (\mathcal{M}^2_{\tilde{f}})_{LR} \\
(\mathcal{M}^2_{\tilde{f}})_{RL}  & (\mathcal{M}^2_{\tilde{f}})_{RR}  \\
\end{pmatrix} \, , \qquad
\tilde{f}=\tilde{u},\tilde{d},\tilde{e} \, \ .
\eeq
The off-diagonal blocks, corresponding to LR/RL mixing,
can be neglected in our model, because the trilinear couplings ($A$-terms)
are negligible.
Hence, the diagonalization of $\mathcal{M}^2_{\tilde{f}}$ splits into the
diagonalization of the two independent LL and RR $3\times 3$ blocks:
\beq
D_{L,\tilde{f}}^2 = (W_L^{\tilde{f}})^\dag
(\mathcal{M}_{\tilde{f}}^2)_{LL} W_L^{\tilde{f}} \, , \qquad
D_{R,\tilde{f}}^2 = (W_R^{\tilde{f}})^\dag
(\mathcal{M}_{\tilde{f}}^2)_{RR} W_R^{\tilde{f}} \, ,
\label{eq:sfermiondiag}
\eeq
in terms of which the quark-squark-gluino mixing matrices are given by
\beq
(Z^f_L)_{ij} = -(U^f_L)^{\dag}_{ik} \, (W_L^{\tilde{f}})_{kj} \, , \qquad
(Z^f_R)_{ij} = (U^f_R)^{\dag}_{ik} \, (W_R^{\tilde{f}})_{kj} \, , \qquad
f = u,d \, ,
\eeq
with $i,j,k=1,2,3$.

Let us now describe the calculation of the FCNC contributions
to the $K-\bar{K}$ mixing in our model.
The squark-gluino box contribution to the neutral kaon mixing can be
parametrized by
\beq
\mathcal{H}= C_1 O_1 + \tilde{C}_1 \tilde{O}_1 + C_4 O_4+ C_5 O_5 \, ,
\eeq
where
\beq
O_1 = \bar{d}_L^\alpha \gamma_\mu s_L^\alpha
  \bar{d}_L^\beta \gamma^\mu s_L^\beta \, ,
\qquad
O_4 = \bar{d}_R^\alpha s_L^\alpha \bar{d}_L^\beta  s_R^\beta \, ,
\qquad
O_5 = \bar{d}_R^\alpha s_L^\beta \bar{d}_L^\beta  s_R^\alpha \,
. \label{operatori}
\eeq
$\tilde{O}_1$ is related to $O_1$ by left and right interchange.
There are generically more operators
(i.e.~$O_2,O_3,\tilde{O}_2,\tilde{O}_3$) which are negligible
here due to the negligible $A$-terms.

The coefficients in front of the operators taken at the superpartner
scale are given by \cite{Hagelin,GGMS,NW}
\begin{align}
C_1 &= \alpha_s^2
\sum_{i,j=1}^3 \left(\frac{11}{36} A_{ij}^{LL} + \frac{1}{9}
B_{ij}^{LL}\right)
(Z^d_L)^*_{1i} (Z^d_L)_{2i} (Z^d_L)^*_{1j} (Z^d_L)_{2j} \, , \non
\tilde{C}_1 &= \alpha_s^2
\sum_{i,j=1}^3 \left(\frac{11}{36} A_{ij}^{RR} + \frac{1}{9} B_{ij}^{RR}\right)
(Z^d_R)^*_{1i} (Z^d_R)_{2i} (Z^d_R)^*_{1j} (Z^d_R)_{2j} \, , \non
C_4 &= \alpha_s^2
\sum_{i,j=1}^3 \left(-\frac{1}{3} A_{ij}^{LR} + \frac{7}{3} B_{ij}^{LR}\right)
(Z^d_L)^*_{1i} (Z^d_L)_{2i} (Z^d_R)^*_{1j} (Z^d_R)_{2j} \, , \non
C_5 &= \alpha_s^2
\sum_{i,j=1}^3 \left(\frac{5}{9} A_{ij}^{LR} + \frac{1}{9} B_{ij}^{LR}\right)
(Z^d_L)^*_{1i} (Z^d_L)_{2i} (Z^d_R)^*_{1j} (Z^d_R)_{2j} \, ,
\label{operatorini}
\end{align}
where the $3\times 3$ matrices $A^{MN},B^{MN}$, $M,N=L,R$, are given
by
\begin{align}
A_{ij}^{MN} &=
\frac{m_{\tilde{g}}^2}{(m_i^2-m_{\tilde{g}}^2)(m_j^2-m_{\tilde{g}}^2)}
+ \frac{m_i^4}{(m_i^2-m_j^2)(m_i^2-m_{\tilde{g}}^2)^2} \log \left(
\frac{m_i^2}{m_{\tilde{g}}^2}\right)
+\frac{m_j^4}{(m_j^2-m_i^2)(m_j^2-m_{\tilde{g}}^2)^2} \log \left(
\frac{m_j^2}{m_{\tilde{g}}^2}\right) \, , \non
B_{ij}^{MN} &=
\frac{m_{\tilde{g}}^2}{(m_i^2-m_{\tilde{g}}^2)(m_j^2-m_{\tilde{g}}^2)}
+\frac{m_i^2 m_{\tilde{g}}^2 }{(m_i^2-m_j^2)(m_i^2-m_{\tilde{g}}^2)^2}
\log \left( \frac{m_i^2}{m_{\tilde{g}}^2}\right)
+\frac{m_j^2 m_{\tilde{g}}^2 }{(m_j^2-m_i^2)(m_j^2-m_{\tilde{g}}^2)^2}
\log \left( \frac{m_j^2}{m_{\tilde{g}}^2}\right) \, , \nonumber
\end{align}
with $m_i\equiv \big(D_{M,\tilde{d}}\big)_{ii}$, $m_j\equiv
\big(D_{N,\tilde{d}}\big)_{jj}$ (see eq.~\eqref{eq:sfermiondiag})
and $m_{\tilde{g}}$ is the gluino mass.
These operators should be evolved by RG
equations down to approximately the hadronic scale $\mu=2 {\rm GeV}$.
This can be done using
the so-called magic numbers; see e.g.~\cite{Ciuchini:1998ix}.
Finally, we can evaluate the contribution to the $K_L-K_S$ mass
difference as\footnote{For the matrix elements of the operators, see
  e.g.~\cite{Bona:2007vi}.}
\beq
\Delta m_K = 2 \, {\rm Re}
\left\langle \left.K^0\right|\mathcal{H}\left|\bar{K}^0\right.\right\rangle \, .
\eeq
Constraints due to $B_d-\bar{B}_d$ and $B_s-\bar{B}_s$ mixing can also be
evaluated using eq.~\eqref{operatorini}, by substituting
$(Z^d_{L,R})_{1i}$ and $(Z^d_{L,R})_{2i}$ with
$(Z^d_{L,R})_{1i}$ and $(Z^d_{L,R})_{3i}$, respectively (for
$B_d-\bar{B}_d$) or with
$(Z^d_{L,R})_{2i}$ and $(Z^d_{L,R})_{3i}$, respectively (for
$B_s-\bar{B}_s$).
The constraints coming from $D-\bar{D}$ mixing are similarly evaluated using
eq.~\eqref{operatorini}, by changing $Z^d$ to $Z^u$.
The magic numbers for $B_q-\bar{B}_q$ mixing are given in
\cite{Becirevic:2001jj}, while those for $D-\bar{D}$ mixing are in
\cite{Bona:2007vi}. The experimental constraints that
we have used are shown in table \ref{tab:FCNClimits}.

Constraints from $b \rightarrow s \gamma$ (due to the gluino-squark loop)
have been computed using
expressions in \cite{Everett:2001yy,Lunghi:2006hc}; these constraints
are linearly enhanced by $\tan\beta$ and are in principle important
especially if there is interference between the new physics and the SM
contributions \cite{Giudice:2008uk}. We found that
these constraints are always satisfied in our model.

\begin{table}[!ht]
\begin{center}
\begin{tabular}{c|r@{}l}
& \multicolumn{2}{c}{Experimental value}  \\   \hline
$\Delta m_{K^0}$ & $3.483 \times$ & $10^{-12}\, {\rm MeV}$  \\
$\Delta m_{D^0}$ & $1.57 \times$ & $10^{-11} \, {\rm MeV}$  \\
$\Delta m_{B^0}$ &  $3.337 \times$ & $ 10^{-10} \, {\rm MeV}$ \\
$\Delta m_{B_s}$ &  $1.163 \times$ & $ 10^{-8} \, {\rm MeV}$ \\
$|\epsilon_K|$ & $2.228\times$ & $ 10^{-3}$\\
$|\epsilon_K^{\rm SM}|$ &  $1.8 \times$ & $ 10^{-3}$\\
$BR(b \rightarrow s \gamma)\phantom{^{\rm SM}}$ & $(3.55\pm 0.24) \times$ & $ 10^{-4}$\\
$BR(b \rightarrow s \gamma)^{\rm SM}$ & $(3.15\pm 0.23) \times$ & $ 10^{-4}$\\
$BR(\mu \rightarrow e \gamma)\phantom{^{\rm SM}}$ & $<2.4 \times$ & $ 10^{-12}$\\
\end{tabular}
\caption{Limits used in the FCNC analyses; $|\epsilon_K|$ is the
  measured value of the kaon $\epsilon$-parameter, while
  $|\epsilon_K^{\rm SM}|$ is a calculation of the Standard Model
  contribution \cite{BG}. The experimental value of $BR(b \rightarrow
  s \gamma)$ is for $E_\gamma>1.6 \, {\rm GeV}$;
  the Standard Model calculation $BR(b \rightarrow s \gamma)^{\rm SM}$
  is taken from \cite{Misiak:2006zs}.}
\label{tab:FCNClimits}
\end{center}
\end{table}

We now return to inspecting the big parameter space of the model discussed
above. We randomly choose the matrices $A,B,C$ of
eq.~\eqref{rotazioni} to find coefficients $\lambda^{u,d}_{ij}$ of
order one, which produce the measured fermion masses
and the CKM matrix elements.
For concreteness, let the $\lambda$s be in a certain
range, say $[0.11,1.1]$. We have done some statistics by finding
random coefficients $\lambda^{u,d}_{ij}$ of the model such that all
the neutral meson oscillation constraints are within the experimental
bounds\footnote{Due to hadronic uncertainties, we impose that
  the new physics contribution to each meson oscillation does not
  exceed the experimental bound on each of them.}.
For simplicity, we have used in the analysis the running fermion
masses at the scale of the top mass (see e.g.~table in 1 in
\cite{Babu}).
By repeating this exercise 5,000 times we get within an accuracy of
roughly $1.4\%$ that $41\%$ of randomly chosen coefficients give rise
to specific model data passing all the above-mentioned flavor
constraints in the case of the spectrum shown in the first column of
table \ref{masses2nodes}, i.e.~in the low-scale mediation example. For
the high-scale mediation example (with masses shown in the second
column of table \ref{masses2nodes}) we find by the same analysis that
$45\%$ of the randomly chosen model points satisfy all the flavor
constraints.
This shows that the model has a large space of parameters which works
out well in terms of the quark masses, the CKM matrix
and flavor constraints.

If an order one complex phase is added to the rotation matrix of the
right-handed squarks, constraints from CP violation are rather
stringent. One constraint is coming from the $\epsilon_K$ parameter
in the neutral kaon system,
\beq
\epsilon_K = \frac{1}{\sqrt{2} \Delta m_K} {\rm Im}
\left.\left\langle K^0 \right|\mathcal{H}\left|\bar{K}^0\right.\right\rangle \, .
\eeq
There are $15$ phases that in principle can contribute to this
observable ($9$ from $Y^d$, $3$ from $(\mathcal{M}_{\tilde{d}}^2)_{LL}$ and
$3$ from $(\mathcal{M}_{\tilde{d}}^2)_{RR}$).
If the masses of the first two generations of squarks are equal,
we can use a symmetry space of dimension $13$ to
set some of the phases to zero. This implies that only $2$
independent physical phases can contribute to $\epsilon_K$.
The SM prediction \cite{BG}, however, is rather firm (see table
\ref{tab:FCNClimits}) and therefore we limit the contribution from new
physics in our model to be within the difference
$|\epsilon_K|-|\epsilon_K^{\rm SM}|$ which
typically means that the two independent complex phases situated in
the right-handed rotation matrix $U_R^d$ should be tuned at the
percent level.\footnote{This observation is consistent with the
  results of \cite{CGK}.}
This is left as a weak point of our model.

Flavor violations in the leptonic sector could be induced by
the non-diagonal couplings in $Y^e$ (see eq.~\eqref{eq:yukawas}).
However, here the situation with $\mu \rightarrow e \gamma$ is rather robust,
due to the degeneracy of the selectron and smuon masses;
using eq.~(20) in \cite{GGMS} as an estimate, a typical example
with order one non-diagonal couplings, gives at worst about $1/30$ of
the current experimental limits.

Above, we presented the results of the analysis done for $\tan\beta=20$.
We have done the statistical analysis also in the case
$\tan\beta=50$ and obtained similar results;
the number of model points satisfying the flavor constraints
increased slightly with respect to that of $\tan\beta=20$.

\section{Higgs mass and naturalness} \label{sec:higgs}

It is well known that there is some tension in the MSSM between the
LEP bounds on the Higgs mass and naturalness.
On the one hand, in gauge-mediated models, a stop with mass of about $1$ TeV or
more is needed for feeding quantum corrections into the Higgs mass in
order to raise it above $114$ GeV. On the other hand, the stop mass
should not be much heavier than the weak scale in order to cut off the
quadratic divergences of top loops.
This tension is enhanced significantly if the Higgs is heavier.
Recently, the ATLAS and CMS collaborations have presented
evidence for a SM-like  Higgs boson with a mass of about
$125$ GeV \cite{atlas,cms}.
If a Higgs boson with such mass will be discovered,
a stop of about $5$ TeV or above would be needed,
leading to about $0.1\%$ fine tuning or worse.

The root of this little hierarchy problem
is due to the fact that in the MSSM the
tree-level mass of the Higgs is bounded by $m_Z$ and hence a big loop
correction is needed.
Interestingly, in our class of models there exists a natural mechanism
for raising the tree-level Higgs mass, because the D-term of the heavy
gauge bosons does not decouple completely in the presence of SUSY
breaking.
The usual MSSM D-terms are modified to \cite{arXiv:hep-ph/0309149}
\beq
V_{D}= \frac{g_2^2 (1+\Delta_2)}{8} \left|H_u^\dagger \sigma^a H_u + H_d^\dagger \sigma^a H_d\right|^2
+\frac{3}{5} \frac{g_1^2 (1+\Delta_1)}{8} \left|H_u^\dagger H_u - H_d^\dagger H_d\right|^2 \, ,
\label{nondecD}
\eeq
where $\sigma^a$ are the Pauli matrices and $\Delta_k$ are given by
\beq
\Delta_k = \frac{\alpha^{-1}_{B_k}}{\alpha^{-1}_{A_k} } \frac{2
  m_{\chi_k}^2}{m_{v_k}^2+ 2 m_{\chi_k}^2} \, , \qquad
m_{\chi_1}^2=\frac{m^2_{\chi_Q}+8 m^2_{\chi_u}+6 m^2_{\chi_e}}{15} \,
, \qquad m_{\chi_2}^2=m^2_{\chi_Q} \, ,
\eeq
where $m^2_{\chi_R}$ are the soft masses of the link fields.
In the presence of $\Delta_{1,2}$, the usual bound $m_{h_0} < m_Z$ at
tree level (which is saturated at large $\tan \beta$) is replaced by
\beq
m_{h_0}^2 < \frac{ \frac{3}{5} g_1^2(1+ \Delta_1) +g_2^2
  (1+\Delta_2)}{2} \, v_{\rm ew}^2 \, ,
\qquad v_{\rm ew} = 174 \, {\rm GeV}\, .
\label{mmvvhh}
\eeq
If $\Delta_{1,2}$ are of order one, this contribution is quite useful
for ameliorating the little hierarchy problem.
This puts an upper bound on $m_{v_k}$, of the order of the link-field soft masses
(which are at most about $10$ TeV). A second requirement is that
$\alpha^{-1}_{B_k}$ is not too small compared to
$\alpha^{-1}_{A_k}$.\footnote{To maximize the effect, we should
  send $g_{A_k}$ to $\infty$ and hence $g_{B_k}$ goes towards the
  effective coupling $g_k$. This would of course take us out of the
  perturbative regime.}
When the link fields $\chi,\tilde{\chi}$ are bifundamentals of
$SU(5)$, this mechanism can raise the Higgs mass rather
effectively. For instance, remaining in the perturbative regime, for
$g_{A_k} \approx g_{B_k}$ it provides a $140$ GeV Higgs with the
stop mass near the TeV \cite{arXiv:1107.1414}.

The main obstacle in realizing this mechanism in a model with $\chi_Q$
link fields is that we generically need $m_{v_k} \approx M$
to be far away from Landau poles, and $M$ should be at least a
few $100$'s of TeV to give the right sparticle masses.
Nevertheless, if we stretch the scales of the problem to the
limit, this is marginally possible in the low-mediation regime;
if the irrelevant operators of the superpotential \eqref{superpote}
have large enough coefficients (of order of a few), and all the
link-field matter has mass of order $m_v\approx 10$ TeV, a Landau pole is
reached near $20 m_v$, which is just enough to accommodate for the messengers.
This corner of parameter space is in a regime where we can only
marginally trust perturbative calculations.
Yet, this regime is also motivated for the following reason.
The condensed values of the various scales in this case,
provide a possible solution to the $\mu$ problem:
following the strategy of \cite{CGK},
we can prohibit a direct $\mu H_u H_d$ term by some symmetry, and
then obtain the $\mu$-term from the operator $\chi\tilde{\chi} H_u
H_d/\Lambda$;
it gives the correct order of magnitude if this irrelevant operator
has an order one coefficient.

\begin{figure}[!ht]
\begin{center}
\includegraphics[width=0.6\linewidth]{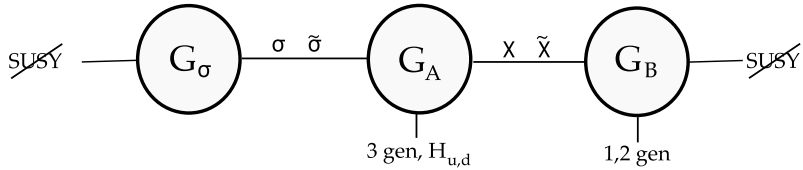}
\caption{A diagram describing a modification of the model
in order to give an extra tree-level contribution to the Higgs mass,
ameliorating the little hierarchy problem.}
\label{fig:quiverfancy}
\end{center}
\end{figure}

A possible alternative is shown in figure \ref{fig:quiverfancy}.
We add an extra gauge group $G_\sigma$, taken to be $SU(2)$,
and an extra set of link fields $\sigma,\tilde{\sigma}$ in the
bifundamental of
$SU(2)_\sigma \times SU(2)_A$.
Suppose now that $\sigma,\tilde{\sigma}$ get a VEV $v_\sigma \approx
10^4$ GeV, e.g.~using irrelevant operators as in the case of
$\chi,\tilde{\chi}$. Assuming that $v_\chi \gg v_\sigma$ and coupling
$G_\sigma$ to the SUSY-breaking messenger, then the link field
$\sigma$ gets a gauge-mediated soft mass $m_\sigma$ providing the
Higgs potential with a contribution as in eq.~\eqref{nondecD}, with
\beq
\Delta_1=0 \, , \qquad
\Delta_2 = \frac{\alpha^{-1}_{\sigma}}{\alpha^{-1}_{A_2} }
\frac{2 m_{\sigma}^2}{m_{v_\sigma}^2+ 2 m_{\sigma}^2} \, ,
\qquad
m_{v_\sigma}^2=2 (g^2_\sigma+g_2^2) v_\sigma^2 \, .
\eeq
Following \cite{arXiv:hep-ph/0309149}, it might even
be possible to obtain unification as well as having a heavy Higgs, by
taking the VEV of $\chi$ to be near the GUT scale. Only the running
of the $SU(2)$ gauge coupling is potentially affected, but the two new
contributions, i.e.~from the extra gauge bosons as well as from the
link fields $\sigma$, cancel out.
Several alternatives, generically not consistent with unification,
could be considered. For instance, we could take $G_\sigma=U(1)$ or
$SU(2) \times U(1)$, as well as various representations
for $\sigma$. Other ways of achieving unification might also be a
possibility here; see e.g.~\cite{hep-th/0108089,hep-ph/0409127}.

One should note that in parameter space, 
the LHC Higgs production cross section could deviate
from that of the Standard Model.
For a recent study of this issue in the MSSM, 
both with and without extra D-terms,
see \cite{Arvanitaki:2011ck}.
Constraints from $b \rightarrow s \gamma$ (due to charged Higgs-top and chargino-stop loops)
can be important and depend on the details of the spectrum in the Higgs sector.

Another way to achieve a heavier Higgs boson
is to couple the Higgs sector to the SUSY-breaking one,
in order to generate a large trilinear $A_t$-term at the messenger scale.
For instance, this is possible if the Higgses
mix with doublet messengers~\cite{Evans:2011bea}.

\section{Three nodes model} \label{sec:3nodes}

\begin{figure}[!ht]
\begin{center}
\includegraphics[width=0.4\linewidth]{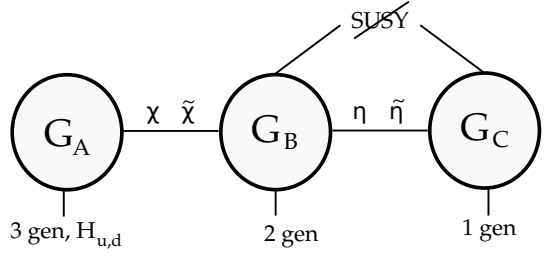}
\caption{A diagram describing the model with $G_A,G_B,G_C$ being gauge
  groups and $\chi,\tilde{\chi},\eta,\tilde{\eta}$ being link fields. }
\label{fig:quiver3nodes}
\end{center}
\end{figure}

To generate a hierarchy also between the first and second
generations of SM fermion masses without tuning any coefficients, we
consider the three-nodes extension shown in
fig.~\ref{fig:quiver3nodes}.
In this example, both the link fields $\chi,\eta$ are taken
in the representations
$\{\chi_Q,\chi_{u^c},\chi_{e^c}\}$, $\{\eta_Q,\eta_{u^c},\eta_{e^c}\}$.

Introducing higher dimension operators, as before, this model gives
rise to Yukawa matrices of the form
\begin{align}
Y_u &=
\begin{pmatrix}
\epsilon_u \epsilon_Q \delta_Q \delta_{u} & \epsilon_u \epsilon_Q
\delta_Q&  \epsilon_Q \delta_Q\\
\epsilon_u \epsilon_Q   \delta_u & \epsilon_u \epsilon_Q  &
\epsilon_Q \\
 \epsilon_u \delta_u  & \epsilon_u & 1
\end{pmatrix} \, , \qquad
Y_d =
\begin{pmatrix}
\epsilon_u \epsilon_Q \delta_Q \delta_{u} & \epsilon_u \epsilon_Q
\delta_Q&  \epsilon_Q \delta_Q\\
\epsilon_u \epsilon_Q \delta_Q^2  & \epsilon_u \epsilon_Q  &
  \epsilon_Q \\
 \epsilon_Q^2   \delta_Q^2  & \epsilon_Q^2 & 1
\end{pmatrix}\, , \\
Y_e &=
\begin{pmatrix}
\epsilon_u \epsilon_Q \delta_Q \delta_{u} &\epsilon_u \epsilon_Q
\delta_Q \delta_{u} & \epsilon_u \epsilon_Q \delta_Q \delta_{u} \\
 \epsilon_u \epsilon_Q  \delta_e & \epsilon_u \epsilon_Q  & \epsilon_u
 \epsilon_Q \\
 \epsilon_e \delta_e  & \epsilon_e & 1  \\
\end{pmatrix} \, , \nonumber
\end{align}
where $\epsilon_R = \langle\chi_R\rangle/\Lambda$ and $\delta_R =
\langle\eta_R \rangle/\Lambda$.
The texture of $Y^u$ is again identical to the
one realized in single-sector SUSY breaking
\cite{hep-ph/9812290,arXiv:0911.2467}; $Y^d$ instead is slightly
different in the lower triangular part.
For generic order one coefficients, $\lambda_{ij}^{u,d,e}$,
these textures predict
\beq
\frac{m_c}{m_t} \sim  \frac{m_s}{m_b}  \sim  \frac{m_\mu}{m_\tau}
  \sim \mathcal{O} (\epsilon_u \epsilon_Q)  \, , \qquad
\frac{m_u}{m_c} \sim   \frac{m_d}{m_s}
\sim  \frac{m_e}{m_\mu}  \sim \mathcal{O} (\delta_u \delta_Q) \, ,
\qquad
\frac{m_t}{m_b} \sim \frac{m_t}{m_\tau} \sim\tan\beta\, .
\eeq

The MSSM gauge couplings $g_{k}$ are given in terms of
the gauge couplings of the unbroken theory by
\beq
\frac{1}{g_{k}^2}= \frac{1}{g_{A_k}^2} + \frac{1}{g_{B_k}^2}
+ \frac{1}{g_{C_k}^2} \, ,
\eeq
where $k=1,2,3$ correspond to the $U(1)$, $SU(2)$ and
$SU(3)$ gauge groups, respectively.
Above the Higgsing scale the gauge $\beta$-function coefficients read
\beq
b_{A_1}=\frac{49}{5},\,\,\, b_{A_2}=15, \,\,\, b_{A_3}=8, \,\,\,
b_{B_1}=\frac{82}{5},\,\,\, b_{B_2}=32,\,\,\, b_{B_3}=23,\,\,\,
b_{C_1}=\frac{46}{5},\,\,\, b_{C_2}=14,\,\,\, b_{C_3}=8\, .
\eeq
%\beq
%b_{A_1}=\frac{49}{5} \, , \qquad b_{A_2}=15 \, , \qquad b_{A_3}=8 \, , \qquad
%b_{B_1}=\frac{82}{5} \, , \qquad b_{B_2}=32 \, , \qquad b_{B_3}=23 \, ,
%\eeq
%\[
%b_{C_1}=\frac{46}{5} \, , \qquad b_{C_2}=14 \, , \qquad b_{C_3}=8 \, .
%\]
Running by a factor of ten in energy decreases $\alpha_{B_2}^{-1}$ by
an amount of about $12$, and $\alpha_{B_3}^{-1}$ by $10$,
$\alpha_{A_2}^{-1},\alpha_{C_2}^{-1}$ by $5$ and
$\alpha_{A_3}^{-1},\alpha_{C_3}^{-1}$ by $3$.
Hence, the running of the gauge couplings is even faster than in
the two-nodes model.
The gauge couplings are running into Landau poles in about a factor
of ten in energy above the Higgsing scale,
and thus the regime $M\gg m_v$ is unattainable.

The main experimental challenge for this model is due to the
constraints from $K-\bar{K}$ and $D-\bar{D}$ oscillations,
which allow just for tiny differences between the first and the second
generations of squark masses.
For this purpose, we introduce an approximate $\mathbb{Z}_2$ symmetry,
forcing the gauge couplings of $G_B$ and $G_C$ to be the same and also
the messengers coupled to the gauge groups have the same mass and are
coupled in the same way to a common SUSY-breaking spurion.
In order not to break this approximate symmetry by the different
running of the couplings $g_{B,C}$, one may consider adding extra
matter to $G_C$ such that the running becomes approximately equal.

\subsection{Sparticle spectrum}

Let us define the quantities $v_{\epsilon k},v_{\delta k}$ in analogy
with the two-nodes case in eq.~\eqref{vevvini}.
The mass-squared matrix of the gauge bosons is given by
\beq
\mathcal{M}_{V_k}^2 = 2
\begin{pmatrix}
g_{A_k}^2 v_{\epsilon k}^2  & - g_{A_k} g_{B_k} v_{\epsilon k}^2 & 0 \\
- g_{A_k} g_{B_k} v_{\epsilon k}^2 & g_{B_k}^2\left(v_{\epsilon k}^2 + v_{\delta k}^2\right) &
  - g_{B_k} g_{C_k} v_{\delta k}^2 \\
0 & - g_{B_k} g_{C_k} v_{\delta k}^2 &   g_{C_k}^2   v_{\delta k}^2
\end{pmatrix} \ .
\eeq
Hence, the masses of the heavy gauge bosons are
\beq
(m_{v_k}^\mp)^2 =a_{k}+ b_{k}
\mp
 \sqrt{ (a_{k}+ b_{k})^2-
4 (g_{A_k}^2 g_{B_k}^2 +g_{B_k}^2 g_{C_k}^2 +g_{A_k}^2 g_{C_k}^2) v_{\epsilon k}^2 v_{\delta k}^2  } \, ,
\eeq
where $a_{k}\equiv (g_{A_k}^2+g_{B_k}^2) v_{\epsilon k}^2$ and
$b_{k}\equiv (g_{B_k}^2+g_{C_k}^2 ) v_{\delta k}^2$.

We can again apply the results of \cite{Auzzi:2011wt}.
Defining the variables
\beq
x=\frac{F}{M^2} \, , \qquad
y_{k}^- =\frac{m_{v_k}^-}{M} \, , \qquad
y_{k}^+= \frac{m_{v_k}^+}{M} \, ,
\eeq
the soft masses of the sfermions are given by eq.~\eqref{sferm2},
where $\mathcal{E}^l$ is different for each generation $l=1,2,3$:
\begin{align}
\mathcal{E}^1 &= n_{C_k} \mathcal{K}
\left(x,y_{ k}^-, y_{ k}^+, 2 \frac{g_{B_k}^2 g_{C_k}^2 v_{\delta k}^2 +
g_{C_k}^2 (g_{A_k}^2+g_{B_k}^2) v_{\epsilon k}^2}{M^2 g_{k}^2} ,\frac{g_{C_k}^2}{g_{k}^2} \right) +
n_{B_k}   \mathcal{K} \left(  x,y_{k}^-, y_{k}^+, 2 \frac{g_{B_k}^2
  g_{C_k}^2 v_{\delta k}^2 }{M^2 g_{k}^2} , 0    \right) \, , \non
\mathcal{E}^2 &= n_{C_k} \mathcal{K} \left(x,y_{k}^-, y_{k}^+, 2 \frac{g_{B_k}^2 g_{C_k}^2 v_{\delta k}^2 }{M^2 g_{k}^2}  ,0  \right) +
n_{k_B}   \mathcal{K} \left(  x,y_{k}^-, y_{k}^+,
2 \frac{g_{A_k}^2 g_{B_k}^2 v_{\epsilon k}^2  +   g_{B_k}^2 g_{C_k}^2
  v_{\delta k}^2  }{M^2 g_{k}^2} , \frac{g_{B_k}^2}{g_{k}^2}
\right) \, , \\
\mathcal{E}^3 &= n_{C_k} \mathcal{K} \left(x,y_{k}^-, y_{k}^+,0,0 \right) +
 n_{B_k}   \mathcal{K} \left(  x,y_{k}^-, y_{k}^+,
2 \frac{ g_{A_k}^2 g_{B_k}^2 v_{\epsilon k}^2  }{M^2 g_{k}^2} , 0
\right) \, , \nonumber
\end{align}
where the function $\mathcal{K}$ is defined in eq.~(4.14) of
\cite{Auzzi:2011wt}, and $n_{B_k}$, $n_{C_k}$ are Dynkin indices of
the messengers coupled to $G_B$ and $G_C$, respectively.
The soft masses of the gauginos are given by eq.~\eqref{mgauginos},
with $n_k=n_{B_k}+n_{C_k}$.
The RG evolution down to the weak scale is done in a similar way to
that of the two-nodes case.

\begin{table*}[!ht]
\begin{center}
\begin{tabular}  {| l | l | l | } \hline
 &  $M=8 \times 10^{5}$  &  $M= 10^{15}$ \\
$\tan\beta = 20$ &   $F/M= 10^5$  &   $F/M=0.8 \times 10^5$ \\
\hline
$(\alpha^{-1}_{B_1,C_1},\alpha^{-1}_{B_2,C_2},\alpha^{-1}_{B_3,C_3})$
& $(15,6.5,2.5)$  & $(5,5,5)$   \\
$(v_{\epsilon Q},v_{\epsilon u},v_{\epsilon e})$   & $(M/3,M/3,M)$  & $(M/3,M/3,M/3)$   \\
$(v_{\delta Q},v_{\delta u},v_{\delta e})$   & $( M, M, M)$  & $( M, M, M)$    \\
$(y_1^-,y_2^-,y_3^-)$   & $(1.9,1.2,1.9)$    & $(1.1,1.4,1.4)$   \\
$(y_1^+,y_2^+,y_3^+)$   & $(3.0,4.9,7.9)$    & $(5.6,5.6,5.6)$   \\
\hline
$m_{\tilde{g}}$    & $1556$ & $1225$    \\
$m_{\tilde{\chi}_0}$   &$(264,502,615,645)$ &$(203,402,645,658)$  \\
$m_{\tilde{\chi}_{\pm}}$    & $(503,644)$  & $(402,658)$    \\
\hline
$(m_{\tilde{u}_L}, m_{\tilde{d}_L},m_{\tilde{u}_R},m_{\tilde{d}_R})$
&$(3592,3593, 3434, 3428)$ &$(2056,2057, 1839, 1754)$   \\
$(m_{\tilde{c}_L}, m_{\tilde{s}_L},m_{\tilde{c}_R},m_{\tilde{s}_R})$
&$(3489, 3490, 3335,3329)$ &$(2009,2011,1799,1718)$   \\
$(m_{\tilde{t}_1},m_{\tilde{t}_2},m_{\tilde{b}_1},m_{\tilde{b}_2})$
& $(1164, 1262, 1237, 1278)$  &  $(811,1016,969,1007)$   \\
$(m_{\tilde{e}_R}, m_{\tilde{e}_L},m_{\tilde{\nu}_e})$
& $(355, 1078, 1074)$  & $(958,1209,1207)$    \\
$(m_{\tilde{\mu}_R}, m_{\tilde{\mu}_L},m_{\tilde{\nu}_\mu})$
& $(337, 1047, 1044)$  & $(928,1176,1172)$    \\
$(m_{\tilde{\tau}_1},m_{\tilde{\tau}_2},m_{\tilde{\nu}_\tau})$ &
 $(114, 258, 221)$ &  $(172,383,368)$ \\
\hline
    $(m_{h_0},m_{H_0},m_{A_0},m_{H_{\pm}})$
      &$(116, 619, 619, 624)$   &$(116, 699, 699, 704)$\\
  \hline
 $\mu$  &$627$  & $658$ \\
 \hline
\end{tabular}
\caption{Sparticle masses in units of GeV in some
  numerical examples (with $n_B=n_C=1$), for three nodes, $\tan \beta
  = 20$, and $\delta_{q,u,e}=0.24$, $ \epsilon_{q,u}=0.08$.
  In the first column, $\epsilon_{e}=3 \epsilon_{q,u}$
  is chosen to avoid a tachyonic stau, while in the second
  column $\epsilon_{e}=\epsilon_{q,u}$.}
\label{masses3nodes}
\end{center}
\end{table*}

In general, even for equal gauge couplings and equal messenger sectors
for $G_{B,C}$, there is a splitting between the
first and second generation sfermion masses, which depends
parametrically on all the gauge couplings as well as the precise
values of $\epsilon_R$ and $\delta_R$. To minimize this
splitting, we have chosen $\delta$ slightly larger than $\epsilon$ by
a factor of $\sim 3$ (see example spectra in table
\ref{masses3nodes}).
In several examples we have observed a tendency of a very light
(or even tachyonic) stau. This problem can be ameliorated by
increasing slightly the VEV of $\chi_{e^c}$.

\subsection{Flavor Constraints}

As already mentioned in the previous subsection, the main problem in this
three-nodes version of our model lies in satisfying the $K-\bar{K}$
and $D-\bar{D}$ mixing constraints. It turns out that the $B_q-\bar{B}_q$
mixing constraints are nearly automatically satisfied. This is due to the
improved pattern giving rise to an extra hierarchy between the first
two generations. $K-\bar{K}$ and $D-\bar{D}$ mixing constraints,
on the other hand, pose a bigger challenge with respect to the two-nodes model,
as the first two generations of squark masses are not degenerate.
To quantify how well the model works, we did the same type of
statistical analysis as in the two-nodes case, viz.~we calculated
5,000 model points of order one coefficients $\lambda^{Q,u,e}$ in the
range $[0.11,1.2]$, and we found that $63\%$ of the points satisfy all
the flavor constraints simultaneously in the case of the low-scale
mediation spectrum shown in the first column of table
\ref{masses3nodes}. In the case of the high-scale mediation spectrum,
shown in the second column of table \ref{masses3nodes}, we found that
$33\%$ of the model points satisfy all the flavor constraints. Hence,
we conclude that even though the model potentially has problems with
$K-\bar{K}$ and $D-\bar{D}$ mixing, a large part of parameter space
satisfies the constraints.

Finally, as in the two nodes model, we have also redone the analysis
with $\tan\beta=50$, and we found the same conclusion, i.e.~the number
of model points satisfying the flavor constraints increased slightly
with respect to that of $\tan\beta=20$.

\section{Discussion} \label{sec:discussion}

In this work, we investigated the possibility that flavor hierarchies
are explained by the pattern of allowed gauge-invariant operators
which follow from a quiver-like, UV completed model.
We found that, both in the two and three-nodes examples,
the quark as well as the lepton masses, and the CKM matrix,
are naturally derived and, moreover,
this is consistent with meson mixing constraints
in a considerable part of the parameter space.

Constraints from CP violation are more stringent; for the
model to be consistent with the bounds on the $\epsilon_K$ parameter,
we need to tune a couple of the complex phases at the
percent level.
It would be interesting to find a mechanism protecting the model,
e.g.~by increasing the mass of the right-handed sbottom
\cite{arXiv:1110.6670}.

The texture of the quarks and SM lepton sectors
in our model is rather good.
On the other hand, relative
neutrino masses and their mixing do not come out at the correct
order of magnitude; it would thus be interesting
to extend the model in this direction as well.
It would also be interesting to explore in detail extensions of
the model which can provide extra contributions to the Higgs mass,
as briefly discussed in section \ref{sec:higgs}.

Finally, our observation that the $\chi_Q$ link field
gives rise to natural textures has a simple reasoning.
Clearly, we need either a $\chi_Q$ or a $\chi_L$
to provide an $SU(2)$-charged link.
Now, it turns out that choosing a $\chi_L$ link field,
the $i,j=1,2$ components of the Yukawa matrices (in
the two-nodes case) are given by a single $\epsilon_L$, to leading order.
Thus, to accommodate for the observed mass hierarchy, we need to take
$\epsilon_L\sim 1/100$. This suppresses the quark mixings.
Another effect is that coupling a $\chi_L$ alone to the MSSM matter
and Higgs cannot generate e.g.~the $3i$
elements of the Yukawa matrices, and hence suppresses the quark
mixing even further. Choosing instead the
$\chi_Q$ link field, we need at least a product of
two link fields to generate a
gauge-invariant coupling of the $i,j=1,2$ MSSM fields with the
Higgs (see eq.~\eqref{qhuxx}).
On the other hand, to couple the third generation to the light ones,
requires a single $\chi_Q$
to provide a gauge-invariant operator.
Consequently, $\epsilon_Q$ needs to be around $1/10$
to reproduce the mass hierarchy and, moreover,
the mixing is automatically sufficiently large,
leading naturally to the measured CKM values.

\subsubsection*{Acknowledgments}

We thank Andrey Katz and Zohar Komargodski for discussions.
This work was supported in part
by the BSF -- American-Israel Bi-National Science Foundation,
and by a center of excellence supported by the Israel Science Foundation
(grant number 1665/10).
SBG is supported by the Golda Meir Foundation Fund.

\end{document}